\documentclass[twoside,fleqn]{article}
\usepackage{espcrc2,psfig,epsfig}

\newcommand{\be}{\begin{equation}}
\newcommand{\ee}{\end{equation}}

\newcommand{\AmS}{{\protect\the\textfont2
  A\kern-.1667em\lower.5ex\hbox{M}\kern-.125emS}}

\title{Ultra High Energy Cosmic Rays from Cosmological Relics.}

\author{V. Berezinsky\thanks{Invited talk at 10th Int. Symp. on Very High 
Energy Cosmic Ray Interactions, LNGS (Italy), July 12-17,1998}\\
INFN, Laboratori Nazionali del Gran Sasso,
I--67010 Assergi (AQ), Italy and Institute for Nuclear Research, Moscow, Russia}
 
\begin{document}

\begin{abstract}

Ultra High Energy Cosmic Rays (UHECR) can be a signal from very early 
(post-infationary) Universe. At this cosmological epoch Topological 
Defects (TD) and long lived superheavy (SH) particles are expected to be 
naturally 
and effectively produced. Both of these relics can produce now 
the particles, such as protons and photons, with energies in a great excess 
of what is observed in UHECR, $E \sim 10^{10}\, - \, 10^{11}~GeV$.
The Topological Defects as the 
UHECR sources are critically reviewed and cosmic necklaces and monopolonia 
are identified as most plausible sources.
The relic superheavy particles and monopolonia are shown to be clustering in 
the halo of our Galaxy 
and their decays produce UHECR without GZK cutoff. The observational 
signatures of both models are discussed. 

\end{abstract}

\maketitle

\section{Introduction}
The observation of cosmic ray particles with energies higher than $10^{11}~GeV$
\cite{EHE,FE}  gives a serious challenge to the known mechanisms of 
acceleration. 
The shock acceleration in different 
astrophysical objects typically gives maximal energy of accelerated protons
less than $(1-3)\cdot 10^{10}~GeV$ \cite{NMA}. The unipolar induction can 
provide the maximal energy $1\cdot 10^{11}~GeV$ only for the extreme values 
of the parameters \cite{BBDGP}. Much attention has recently been given to 
acceleration by ultrarelativistic shocks \cite{Vie},\cite{Wax}. The
particles here can gain a tremendous increase in energy,
equal to $\Gamma^2$, at a single reflection, 
where $\Gamma$ is the Lorentz factor of the shock.
However, it is 
 known (see e.g.  the simulation 
for pulsar relativistic wind in \cite{Hosh}) that particles entering 
the shock region are captured there or at least have a small probability 
to escape (see discussion relevant for UHECR in ref.\cite{B99} 

{\em Topological defects} (for a review see \cite{Book}) can naturally 
produce particles of ultrahigh energies (UHE). The pioneering observation 
of this possibility was made by Hill, Schramm and Walker \cite{HS} (for 
a general analysis of TD as UHE CR sources see \cite {BHSS} and for a 
review \cite{Sigl}).

In many cases TD become unstable and decompose to constituent fields, 
superheavy gauge and Higgs bosons (X-particles), which then decay 
producing UHECR. It could happen, for example, when two segments of 
ordinary string, or monopole and antimonopole touch each other, when 
electrical current in superconducting string reaches the critical value
and in some other cases.

In most cases the problem with UHECR from TD is not the maximal energy, 
but the fluxes.  One very general reason 
for the low fluxes consists in the large distance between TD. A dimension
scale for this distance is the Hubble distance $H_0^{-1}$. However, in some 
rather exceptional cases this  scale is multiplied to a small 
dimensionless value $r$. If a distance between TD is larger than 
 UHE proton attenuation length (due to the GZK effect \cite{GZK}), then 
the UHE flux has an exponential cutoff.

The other general restriction to the flux of UHECR comes from the observed 
by EGRET extragalactic flux of gamma-ray radiation at energies between 
$10~MeV$ and $100~GeV$. UHE particles such as photons and electrons produce 
e-m cascade in the Universe, colliding with photons of background 
radiation (most notably microwave and radio background). Low energy part 
of this cascade extends down to the EGRET measurements. The cascade  
flux is below the EGRET limit if the energy density of the cascade photons
$\omega_{cas}<2\cdot 10^{-6}~eV/cm^3$. On the other hand $\omega_{cas}$ 
can be readily evaluated for TD from the total energy release assuming that 
half of it transferred to e-m cascade. Since, on the other hand the same 
energy release determines the UHECR flux, the latter is limited by 
the value of $\omega_{cas}$ given above.

{\em Relic SH particles} have important property of clusterization in 
the Galactic halo and thus UHECR produced at their decays do not have 
the GZK cutoff \cite{BKV}. The same property have monopolonium \cite{BBV}.
The relic SH particles produce UHECR due to their decays. Their lifetime
must be larger (or much larger) than the age of the Universe. For particles 
heavier
than $10^{13} - 10^{14}~GeV$ this is very restrictive condition. 
Even dimension-5 gravitational interaction makes the lifetime of such 
massive particles very short. A particle must be protected from this 
fast decay by some symmetry which is broken extremely weakly, for example 
by instanton effects \cite{KR} or worm hole effects \cite{BKV}.

{\em Production} of both relic SH particles and TD naturally occurs 
in the end of inflation. Decays of inflatons most probably 
start in the regime of  broad parametric resonance \cite{KLS}. It is 
accompanied by massive production of particles not in thermal equilibrium.
At this stage ("preheating") TD can be born due to phase transitions. 
SH particles are produced in the varying gravitational field of 
the oscillating inflaton and can have now critical or subcritical 
density. In the end of preheating phase the thermalization of produced 
particles occurs, and the Universe enters the stage of reheating, when the 
temperature is expected to be large, $T_r \sim \sqrt{\Gamma M_p}$, where 
$M_p$ is the Planck mass and $\Gamma$ is the width of the inflaton decay in 
the regime of the broad parametric resonance. Due to large $\Gamma$ the 
reheating temperature is expected to be as high as $T_r \sim 10^{11} - 10^{12}~
GeV$ or even higher. The superheavy relic particles can be born also at 
the reheating phase.

{\em Spectrum} of UHE particles produced at the decays of a relic SH 
particles or of heavy X particles to which TD decompose, is 
basically the energy spectrum of QCD cascade. This spectrum, in contrast to 
the case of accelerated particles, is essentially non-powerlaw. It 
has the Gaussian peak, the maximum of which determines the multiplicity
\cite{QCD}. For the large masses at interest the supesymmetric effects 
become important; they considerably change the QCD 
spectrum \cite{SUSY-QCD}.
The generic feature of decays of SH particles is the dominance of 
pion production over baryon production. It results in the dominance of 
UHE photons over nucleons at production, roughly as $\gamma/N \sim 10$.

{\em Observational signature} of TD is the presence of UHE photons in 
the primary radiation. At some energies this effect is suppressed by 
strong absorption on radio background \cite{radio}. For the discussion and 
references see \cite{BBV}. The GZK cutoff is present, but 
it is weaker than in case of accelerator sources, due to 
the shape of QCD energy spectrum if the space distribution of the sources is 
the same (e.g. for necklaces). 

In case of relic SH particles and monopolonia (both of them are concentrated 
in the halo) the signature is 
dominance {\em at observation} of UHE photons over nucleons. Another 
signature is anisotropy caused by asymmetric position of the sun in the halo.

\section{Topological Defects}

The following TD have been discussed as potential sources of UHE particles:
superconducting strings \cite{HSW}, ordinary strings \cite{BR}, 
including the cusp radiation \cite{Bran}, networks of monopoles connected by 
strings \cite{BMV}, necklaces \cite{BV}, magnetic monopoles, or more 
precisely bound monopole-antimonopole pairs (monopolonium \cite{Hill,BS}),
and vortons. Monopolonia and vortons are clustering in the Galactic halo,
and UHECR production is thus similar to the case of relic SH particles 
considered in the next section.

(i) {\em Superconducting strings}\\
As was first noted by Witten\cite{Witten}, in a wide class of elementary 
particle models, strings behave like superconducting wires. Moving through 
cosmic magnetic fields, such strings develop electric currents.
Superconducting strings produce X particles when the electric current
in the strings reaches the critical value. In some scenarios,
e.g. \cite{OTW}  where the current is induced by primordial magnetic field,
the critical current produces strong magnetic field, in which all high 
energy particles degrade catastrophically in energy \cite{BeRu}. 
However, for {\em ac} currents there are portions of the string with large 
electric charge and small current. High energy particles can escape from
there.

Large {\it ac} currents can be induced in string loops as they oscillate in
 galactic or extragalactic magnetic fields. Even if the string current
 is typically well below critical, super-critical currents can be
 reached in the vicinity of cusps, where the string shrinks by a large
 factor and density of charge carriers is greatly enhanced. In this
 case, X particles are emitted with large Lorentz factors.

Loops can also acquire {\it dc} currents at the time of formation, when they
are chopped off the infinite strings. As the loops lose their energy
by gravitational radiation, they shrink, the {\it dc} currents grow, and
eventually become overcritical. There could be a variety of
astrophysical mechanisms for
excitation of the electric current in superconducting strings, but for 
all mechanisms considered so far the flux of 
UHE particles is smaller than the observed flux \cite{BeVi}.  However,
the number of possibilities to be explored here is very large, and
more work is needed to reach a definitive conclusion.

(ii) {\em Ordinary strings}\\
There are several mechanisms by which ordinary strings can produce UHE 
particles.

For a special choice of initial conditions, an ordinary  loop can collapse to a
double line, releasing its total energy in the form of X-particles\cite{BR}. 
However, the probability of this mode of collapse is
extremely small, and its contribution to the overall flux of UHE
particles is negligible.

String loops can also 
produce X-particles when they self-intersect (e.g. \cite{Shell}).
Each intersection, however, gives only a few
particles, and the corresponding flux is very small \cite{GK}. 

Superheavy particles with large Lorentz factors can be produced in 
the annihilation of cusps, when the two cusp segments overlap \cite{Bran}.  
The energy released in a single cusp event can be quite large, but
again, the resulting flux of UHE particles is too small to account for
the observations \cite {Bhat89,GK}.

It has been recently argued \cite{Vincent} that long
strings lose most of
their energy not by production of closed loops, as it is generally
believed, but by direct emission of heavy X-particles.
If correct, this claim will change dramatically 
the standard picture of string evolution. It has been also
suggested that the decay products of particles produced in this
way can explain the observed flux of UHECR \cite{Vincent,ViHiSa}. 
However, as it is argued in ref \cite{BBV}, numerical simulations described in
\cite{Vincent} allow an alternative interpretation not connected with 
UHE particle production.  

But even if the conclusions of \cite{Vincent} were correct, the
particle production mechanism suggested in that paper cannot explain
the observed flux of UHE particles. If particles are emitted directly
from long strings, then the distance between UHE particle sources $D$ is
of the order of the Hubble distance $H_0^{-1}$, $D \sim H_0^{-1} \gg R_p$, 
where $R_p$ is the proton attenuation length  in the microwave background 
radiation. In this case UHECR flux has an exponential cutoff at energy 
$E \sim 3\cdot 10^{10}~GeV$. In the case of accidental proximity of a
string to the observer, the flux is strongly anisotropic. A fine-tuning 
in the position of the observer is needed to reconcile both 
requirements.

(iii){\em Network of  monopoles connected by strings}.\\
The sequence of phase transitions
\begin{equation}
G\to H\times U(1)\to H\times Z_N
\label{eq:symm}
\end{equation}
 results in the formation of monopole-string networks in which each monopole 
is attached to N strings. Most of the monopoles and most of the strings belong 
to one infinite network. The evolution of networks is expected to be 
scale-invariant with a characteristic distance between monopoles 
$d=\kappa t$, where $t$ is the age of Universe and $\kappa=const$. 
The production of UHE particles are considered in \cite{BMV}. Each 
string attached 
to a monopole pulls it with a force equal to the string tension, $\mu \sim 
\eta_s^2$, where $\eta_s$ is the symmetry breaking vev of strings. Then
monopoles have a typical acceleration $a\sim \mu/m$, energy $E \sim \mu d$ 
and Lorentz factor $\Gamma_m \sim \mu d/m $, where $m$ is the mass of the 
monopole. Monopole moving with acceleration can, in principle, radiate  
gauge quanta, such as photons, gluons and weak gauge bosons, if the
mass of gauge quantum (or the virtuality $Q^2$ in the case of gluon) is
smaller than the monopole acceleration. The typical energy of radiated quanta 
in this case is $\epsilon \sim \Gamma_M a$. This energy can be much higher 
than what 
is observed in UHECR. However, the produced flux (see \cite{BBV}) is much 
smaller than the observed one. 

(iv){\em Vortons}.\\
Vortons are charge and current carrying loops of superconducting
string stabilized by their angular momentum \cite{dash}.  Although
classically stable, vortons decay by gradually losing charge carriers
through quantum tunneling.  Their lifetime, however, can be greater
than the present age of the universe, in which case the escaping
$X$-particles will produce a flux of cosmic rays.  The $X$-particle
mass is set by the symmetry breaking scale $\eta_X$ of string 
superconductivity.  

The number density of vortons formed in the early universe is rather
uncertain.  According to the analysis in Ref.\cite{BCDT}, vortons are
overproduced in models with $\eta_X > 10^9 GeV$, so all such models
have to be ruled out.  In that case, vortons cannot contribute to the
flux of UHECR.  However, an alternative analysis \cite{mash} suggests
that the excluded range is $10^9 GeV <\eta_X < 10^{12}GeV$, while for
$\eta_X \gg 10^{12}GeV$ vorton formation is strongly suppressed.  This
allows a window for potentially interesting vorton densities
with\footnote{These numbers assume that strings are formed in a
first-order phase transition and that $\eta_X$ is comparable to the
string symmetry breaking scale $\eta_s$.  For a second-order phase
transition, the forbidden range widens and the allowed window moves
towards higher energies \cite{mash}.}
$\eta_X \sim 10^{12}-10^{13}GeV$.  

Like monopolonia and SH relic particles, vortons are  clustering in the
Galactic halo and UHECR production and spectra are similar in these three 
cases. 

(iv){\em Necklaces}.\\
Necklaces are hybrid TD corresponding to the case $N=2$ in 
Eq.(\ref{eq:symm}), i.e. to the case when each monopole is attached to two
strings.  This system resembles ``ordinary'' cosmic strings,
except the strings look like necklaces with monopoles playing the role
of beads. The evolution of necklaces depends strongly on the parameter
\begin{equation}
r=m/\mu d,
\end{equation}
where $d$ is the average separation between monopoles and antimonopoles
along  the strings.
As it is argued in Ref. \cite{BV}, necklaces might evolve to  
configurations with $r\gg 1$, though numerical simulations are needed to 
confirm this conclusion.  
Monopoles and antimonopoles trapped in the necklaces
inevitably  annihilate in the end, producing first the heavy  Higgs and 
gauge bosons ($X$-particles) and then hadrons.
The rate of $X$-particle production can be estimated as \cite{BV} 
\begin{equation}
\dot{n}_X \sim \frac{r^2\mu}{t^3m_X}.
\label{eq:xrate}
\end{equation}

Restriction due to e-m cascade radiation demands the cascade energy density 
$\omega_{cas} \leq 2\cdot 10^{-6}~eV/cm^3$. The cascade energy density 
produced by necklaces can be calculated as
\begin{equation}
\omega_{cas}=
\frac{1}{2}f_{\pi}r^2\mu \int_0 ^{t_0}\frac{dt}{t^3}
\frac{1}{(1+z)^4}=\frac{3}{4}f_{\pi}r^2\frac{\mu}{t_0^2},
\label{eq:n-cas}
\end{equation}
where $f_{\pi}\approx 0.5$ is a fraction of total energy release 
transferred to the cascade.
The separation between necklaces is given by \cite{BV} 
$D \sim r^{-1/2}t_0$ for large $r$. Since $r^2\mu$ is limited by cascade 
radiation, Eq.(\ref{eq:n-cas}), one can obtain a lower limit on the 
separation $D$ between necklaces as
\begin{equation}
D \sim \left( \frac{3f_{\pi}\mu}{4t_0^2\omega_{cas}}\right)^{1/4}t_0
>10(\mu/10^6~GeV^2)^{1/4}~kpc,
\label{eq:xi}
\end{equation}

Thus, necklaces can give a realistic example of the case when separation 
between sources is small and the Universe can be assumed  uniformly filled by 
the sources. 

The fluxes of UHE protons and photons are shown in Fig.1 according to 
calculations of Ref.\cite{BBV}.
Due to absorption of UHE photons the 
proton-induced EAS from necklaces strongly dominate over those induced by 
photons at all 
energies except $E> 3\cdot 10^{11}~GeV$ (see Fig.1), where photon-induced 
showers can comprise an appreciable fraction of the total rate.
The dashed,
dotted and solid lines in Fig.1 correspond to the masses of X-particles 
$10^{14}~GeV,\;\; 10^{15}~GeV$ and $10^{16}~GeV$, respectively. The values 
of $r^2\mu$ used to fit these curves to the data are 
$7.1\cdot 10^{27}~GeV^2,\;\;6.0\cdot 10^{27}~GeV^2$ and 
$6.3\cdot 10^{27}~GeV^2$, respectively. They correspond to the cascade 
density $\omega_{cas}$ equal to 
$1.5\cdot 10^{-6}~eV/cm^3,\;\;
1.2\cdot 10^{-6}~eV/cm^3$ and $1.3 \cdot 10^{-6}~eV/cm^3$, respectively, all 
less than the allowed cascade energy density for which we adopt the 
conservative value $\omega_{cas}=2\cdot 10^{-6}~eV/cm^3$.

For energy lower than $1\cdot 10^{10}~GeV$, the presence of 
another component with a cutoff at $E\sim 1\cdot 10^{10}~GeV$ is assumed. It 
can be 
generated, for example, by jets from AGN \cite{Bier}, which naturally 
have a cutoff at this energy. \\
\vspace{-1cm}
\begin{figure}[h]
\epsfxsize=8truecm 
\centerline{\epsffile{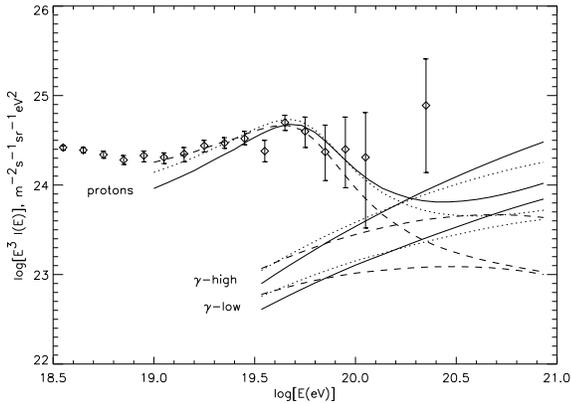}}
\vspace{-1cm}
\caption{Proton and gamma-ray fluxes from necklaces. High ($\gamma$-high)
and low ($\gamma$-low) photon fluxes correspond to two extreme cases of 
gamma-ray absorption. The fluxes are given for $m_X=1\cdot 10^{14}~GeV$ 
(dashed lines), $m_X=1\cdot 10^{15}~GeV$ (dotted lines) and 
$m_X=1\cdot 10^{16}~GeV$ (solid lines).}
\end{figure}

{\section{Relic Superheavy Particles}

{\em Production} of relic SH particles  occurs in time varying gravitational 
field of oscillating inflaton \cite{Kolb,KuTk}. SH particle must be lighter 
than inflaton, otherwise the relic density of SH particles is exponentially 
suppressed. Since inflaton has to be lighter than $10^{13}~GeV$ to produce 
the required spectrum of primordial density fluctuations, this scale gives 
the upper limit to the mass of SH relic particle. In this scenario SH 
particles can provide the critical density of the Universe.

Several other plausible mechanisms were identified in \cite{BKV}, including   
thermal production at the reheating stage, production through the decay of 
inflaton field at the end of the preheating phase 
and through the decay of hybrid topological 
defects, such as monopoles connected by strings or walls bounded by
strings.  

We shall start our short description with the non-equilibrium thermal 
production.

For the thermal production, temperatures comparable to $m_X$ are needed. 
In the case of a heavy decaying gravitino,
the reheating temperature $T_R$ (which is the highest temperature 
relevant for the considered  problem)            
is severely limited to value below $10^8- 10^{10}$~GeV, depending 
on the gravitino mass (see Ref. \cite{ellis} and references therein).  
On the other hand, 
in models with dynamically broken supersymmetry, the lightest 
supersymmetric particle is the gravitino. Gravitinos with mass 
$m_{3/2} \leq 1$~keV  interact relatively strongly with the thermal bath,
thus decoupling relatively late, and it can be the CDM particle \cite{grav}. 
In this scenario all phenomenological
constraints on $T_R$ (including the decay of the second 
lightest supersymmetric particle) disappear and one can assume
$T_R \sim 10^{11} - 10^{12}$~GeV. In this 
range of temperatures, SH particles are not in thermal equilibrium.
If $T_R < m_X$, the density  $n_X$ of $X$-particles produced during the 
reheating phase at time $t_R$ due to $a+\bar{a} \to X+\bar{X}$ is easily 
estimated as
\be
n_X(t_R) \sim N_a n_a^2 \sigma_X t_R \exp(-2m_X/T_R),
\label{eq:dens}
\ee 
where $N_a$ is the number of flavors which participate in the production of 
X-particles, $n_a$ is the density of $a$-particles and $\sigma_X$ is 
the production cross-section. The density of $X$-particles at the
present epoch can be found by the standard procedure of calculating
the ratio $n_X/s$, where 
$s$ is the entropy density. Then for $m_X = 1\cdot 10^{13}$~GeV
and $\xi_X$ in the wide range of values $10^{-8} - 10^{-4}$, the required
reheating temperature is $T_R \sim 3\cdot 10^{11}$~GeV.

In the second scenario mentioned above, non-equilibrium inflaton decay,
$X$-particles are usually overproduced and a second period of 
inflation is needed 
to suppress their density.

Finally, $X$-particles could be produced by TD such as strings or textures. 
Particle production occurs at string intersections or in collapsing texture 
knots. The evolution of defects is scale invariant, and roughly a constant 
number of particles $\nu$ is produced per horizon volume $t^3$ per Hubble 
time $t$. ($\nu \sim 1$ for textures and $\nu \gg 1$ for strings.) The main 
contribution to to the X-particle density is given by the earliest epoch,
soon after defect formation, and we find 
$\xi_X \sim 10^{-6} \nu (m_X/10^{13}~GeV)(T_f/10^{10}~GeV)^3$, where 
$T_f$ is the defect formation temperature. Defects of energy scale
$\eta \geq m_X$ could be formed at a phase transition at or slightly 
before the end of inflation. In the former case, $T_f \sim T_R$ , while in 
the latter case defects should be considered as "formed" when their typical 
separation becomes smaller than $t$ (hence $T_f < T_R$).  

X-particles can also be produced 
by hybrid topological defects: monopoles connected by strings or walls 
bound by strings. The required values of $n_X/s$ can be obtained for a wide 
range of defect parameters.

{\em Lifetime} of SH particle has to be larger (or much larger) than age of 
the Universe. Even gravitational interactions, if unsuppressed, make the 
lifetime of $X$-particle with mass $m_X \sim 10^{13} - 10^{14}~GeV$ much 
shorter. Some (weakly broken) symmetry is needed to protect X-particle 
from the fast decay. Such symmetries indeed exist, e.g. discrete gauge 
symmetries. If such symmetry is very weakly broken by e.g. wormhole 
effects \cite{BKV} or decay is caused by instanton effects \cite{KR},
X-particle can have the desired lifetime. The detailed analysis of the 
gauge discrete symmetries was recently performed in \cite{Yanagida}. There 
were found the cases when allowed non-renormalizable operators for a decay 
of X-particle are suppressed by high power of the Planck mass. In this case 
the lifetime of X-particle can be larger than the age of the Universe. 

The realistic example of long-lived SH particle, the crypton, is given in 
\cite{Benakli}. Like in the case above, decay of crypton is suppressed by 
high power of the Planck mass.

{\em Energy spectrum} of decaying particles from DM halo has no GZK cutoff 
\cite{BKV} and photons dominate in the flux. The energy spectrum was 
calculated using QCD Monte Carlo simulation "Herwig" \cite{Sarkar} and 
as limiting QCD spectrum with supersymmetric particles taken into account 
\cite{BK}. The spectrum, as calculated in \cite{BBV}, is shown in Fig.2.
\begin{figure}[h]
\epsfxsize=8truecm 
\centerline{\epsffile{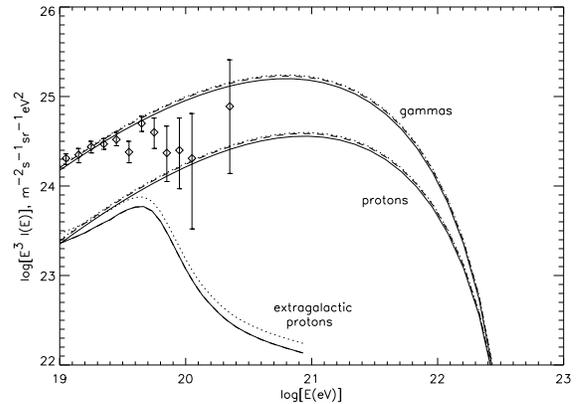}}
\vspace{-1cm}
\caption{Predicted fluxes from relic SH particles 
($m_X=1\cdot 10^{14}~GeV$) or from monopolonia producing X-particles with 
the same masses: nucleons from the halo (curves labelled as "protons"), 
gamma-rays from halo (curves labelled "gammas") and extragalactic protons.
The solid, dotted and dashed curves correspond do different models of 
DM distribution in the halo.}
\end{figure}

\section{Signatures}

In contrast to acceleration astrophysical sources, TD 
production spectrum is enhanced by UHE photons. Though UHE photons are 
absorbed stronger than UHE protons (antiprotons), the photons can 
dominate at some energies or at least $\gamma/p$ ratio in case of TD 
is much larger than in case of acceleration sources \cite{Aha,BBV}.
This signature can be discussed quantitatively for necklaces, probably 
the only extragalactic TD which satisfy the observational constraints.
At large value of $r=m/\mu d >10^7$ necklaces have a small separation 
$D < R_{\gamma}$, where $R_{\gamma}$ is an absorption length for UHE 
photons. They are characterized by a small fraction of photon-induced 
EAS at energies $10^{10} - 10^{11}~GeV$. However, this fraction increases 
with energy and becomes considerable at the highest energies.

Relic SH particles, as well as monopolonia and vortons, have the enhanced 
density in the Galactic halo. The signature of this relics is absence 
of the GZK cutoff, dominance of UHE gamma radiation at observation and 
anisotropy due to non-central position of the Sun in the DM halo. 

{\em Anisotropy} is the strongest signature of the DM halo model. It is
most noticeable as the difference in fluxes between 
directions to Galactic Center and Anticenter. Since  
Galactic Center is not observed by any of existing now detectors,
anisotropy for them is less pronounced, but can be detected if the halo
component becomes dominant at $E \sim (1-3)\cdot 10^{19}~eV$. In case 
the halo component is responsible only for the events at 
$E\geq 1\cdot 10^{20}~eV$ as recent AGASA data suggest,
statistics is too small for the predicted anisotropy and this problem 
will be left for the Auger detector in the southern hemisphere 
\cite{Cronin}.   

{\em UHE photons} as primaries can be also tested by the existing detectors
.   
The search for photon induced showers is not an easy experimental task.
It is known (see e.g. Ref.\cite{AK}) that the muon content of the  
photon-induced showers at very high energies 
is very similar to that in proton-induced showers. However,
some  difference in the muon content between these two 
cases is expected and may be used to distinguish between them
observationally. 

Fly's Eye detector is the most effective one in distinguishing between the 
photon 
and proton induced showers. This detector is able to reconstruct the 
development of the shower in the atmosphere \cite{FE,FE1}, which is 
different for photon and proton induced showers. The analysis \cite{Halzen} 
of the highest energy shower $E \sim 3\cdot 10^{20}~eV$ detected by Fly's Eye 
detector  disfavors its photon production. The future HiRes detector 
\cite{HiRes} will reliably distinguish the photon and proton induced showers
.

The Landau-Pomeranchuk-Migdal (LPM) effect \cite{LPM} and the absorption of 
photons in the geomagnetic field are two other 
important phenomena which affect the detection of UHE photons 
\cite{AK,Kasa}; (see \cite{ps} for a recent discussion). 
The LPM effect reduces the cross-sections 
of electromagnetic interactions at very high energies. However, if a
primary photon approaches the Earth in a direction characterized by a large
perpendicular component of the geomagnetic field, the photon likely decays
into electron and positron \cite{AK,Kasa}. Each of them emits 
a synchrotron photon,
and as a result a bunch of photons strikes the Earth atmosphere. The LPM 
effect, which strongly depends on energy, is thus suppressed. If, on the other 
hand,
a photon moves along the magnetic field, it does not decay, and LPM effect 
makes shower development in the atmosphere very slow. At extremely 
high energies the maximum of the showers can be so close to the Earth surface 
that it becomes "unobservable" \cite{ps}.

\section{Conclusions}

Topological Defects and relic quasistable SH particles are effectively produced
in the post-inflationary Universe and can produce now UHE
particles (photons and (anti)nucleons) with energies higher than observed 
now in UHECR. 

The fluxes from most known TD are too small to explain the observations. 
The plausible candidates are necklaces and monopolonia (the latter by 
observational properties are similar to relic SH particles). The fluxes 
from extragalactic TD are restricted by e-m cascade radiation. The energy
spectrum of UHE (anti)protons from TD have less pronounced GZK cutoff than 
from acceleration sources, because of the QCD production spectrum, which 
is much different from the power-law energy spectrum. The signature of 
extragalactic TD is presence of UHE photons in the primary radiation. 
Absorption of UHE photons on radiobackground considerably diminishes 
the fraction of photon induced showers at observation.

Relic SH particles (and monopolonia) are concentrated in the Galactic halo
and their energy spectrum does not exhibit the GZK cutoff. UHE photon flux 
is $\sim 10$ times higher than that of protons. Detectable anisotropy is 
expected, especially the difference of fluxes between Galactic Center and 
Anticenter.

Therefore, both sources, TD and relic SH particles, have very distinct 
experimental predictions, which can be tested with help of present 
detectors, but most reliably - with the future detectors, such as 
the Auger detector in the south hemisphere \cite{Cronin} and 
HiRes\cite{HiRes}.\\*[3mm]

This paper is based on the talk given at 10th Int. Symposium on Very High 
Energy Cosmic Ray Interaction, July 12-17,1998. The new publications 
which appeared since that time is not included in this review. Most important 
of them is the new data of the 
AGASA detector at energies higher than $1\cdot 10^{20}~eV$. As shown in 
Fig.2 \cite{AGASA} they can be 
interpreted as existence of two components of UHECR, one with the GZK 
cutoff and another - without it and extending to energies 
$(2-3)\cdot 10^{20}~eV$. In this case the DM halo component (relic SH 
particles) have to be responsible only for $E> 1\cdot 10^{20}~eV$ part of 
the spectrum.\\*[2mm]
{\bf Acknowledgements}

I am grateful to my collaborators P.Blasi, M.Kachelriess and A.Vilenkin 
for the discussions and help. M.Hillas, P.Sokolsky and A.Watson are 
thanked for stimulating conversations and correspondence.

\end{document}